\documentclass[aps,pra,floatfix,superscriptaddress,showpacs,twocolumn]{revtex4}
\usepackage{amsmath}
\usepackage{graphicx}
\usepackage{bm}

\def\mum{\mbox{ $\mu$m}}

\def\sqcm{\mbox{ cm$^2$}}

\def\invcucm{\mbox{ cm$^{-3}$}}

\def\ms{\mbox{ ms}}

\def\ea0{\mbox{ $ea_0$}}

\def\K{\mbox{ K}}

\begin{document}
\title{Production of long-lived atomic vapor inside high-density buffer gas}
\author{A.~O.~Sushkov}
\email{alexsushkov@berkeley.edu} \affiliation{Department of Physics,
University of California at Berkeley, Berkeley, California
94720-7300}
\author{D.~Budker}
\affiliation{Department of Physics, University of California at
Berkeley, Berkeley, California 94720-7300} \affiliation{Nuclear
Science Division, Lawrence Berkeley National Laboratory, Berkeley,
California 94720}
\date{\today}
\begin{abstract}
Atomic vapor of four different paramagnetic species: gold, silver,
lithium, and rubidium, is produced and studied inside several buffer
gases: helium, nitrogen, neon, and argon. The paramagnetic atoms are
injected into the buffer gas using laser ablation. Wires with
diameters 25~$\mu$m, 50~$\mu$m, and 100~$\mu$m are used as ablation
targets for gold and silver, bulk targets are used for lithium and
rubidium. The buffer gas cools and confines the ablated atoms,
slowing down their transport to the cell walls. Buffer gas
temperatures between 20~K and 295~K, and densities between
$10^{16}$~cm$^{-3}$ and $2\times10^{19}$~cm$^{-3}$ are explored.
Peak paramagnetic atom densities of $10^{11}$~cm$^{-3}$ are
routinely achieved. The longest observed paramagnetic vapor density
decay times are 110~ms for silver at 20~K and 4~ms for lithium at
32~K. The candidates for the principal paramagnetic-atom loss
mechanism are impurities in the buffer gas, dimer formation and atom
loss on sputtered clusters.
\end{abstract}
\pacs{32.10.-f, 34.50.-s, 33.57.+c, 37.20.+j}



\maketitle

\section{Introduction}

The ability to achieve long-lived coherences in an ensemble of atoms
or molecules is at the core of many of today's atomic physics
experiments. Such experiments are at the forefront of both
technological applications, such as atomic
magnetometers~\cite{Budker2002a,Budker2007}, spin-exchange optical
pumping of noble-gas nuclei~\cite{Happer1972}, and quantum
computation~\cite{Bouwmeester2000}, and more fundamental research,
such as tests of Lorentz invariance~\cite{Berglund1995,Kornack2002}
and searches for violation of the discrete symmetries of
nature~\cite{Khriplovich1997}. In room-temperature experiments
involving ground-state electron-spin coherences, alkali atoms are
most often used inside sealed vapor cells, which are either coated
with an anti-relaxation material, such as
paraffin~\cite{Bouchiat1966,Alexandrov1996}, or filled with a buffer
gas, such as helium~\cite{Happer1972}. An alkali atom inside such a
cell is prepared in a coherent state by optical pumping, the
coherences here are between ground-state magnetic sublevels. In an
anti-relaxation coated cell, it can then experience thousands of
velocity-changing collisions with the cell walls before de-cohering.
Coherence times of 500~ms have been demonstrated in such cells, see,
for example, Ref.~\cite{Budker1998}. In a buffer-gas filled cell,
depolarization due to cell-wall collisions is suppressed, since the
alkali atoms no longer travel ballistically, but have to diffuse
inside the buffer gas, whose density ranges typically from
$10^{16}$~cm$^{-3}$ up to $10^{19}$~cm$^{-3}$. However, as the
buffer-gas density is increased, slowing down alkali diffusion,
collisional de-coherence starts to dominate. This refers to
spin-relaxation of the alkali atom in a collision with a buffer-gas
atom due to spin-orbit interaction~\cite{Bernheim1962}. Coherence
times of 20~ms have been achieved in buffer-gas filled
cells~\cite{Brandt1997,Erhard2001}. Collisional relaxation
cross-sections of the alkali atoms have been studied at room
temperature and above, but not for lower temperatures. The reason
for this is the insufficient saturated vapor pressure of the alkalis
below room temperature, which makes experiments with sealed vapor
cells extremely hard.

\begin{figure}[h]
    \includegraphics[width=\columnwidth]{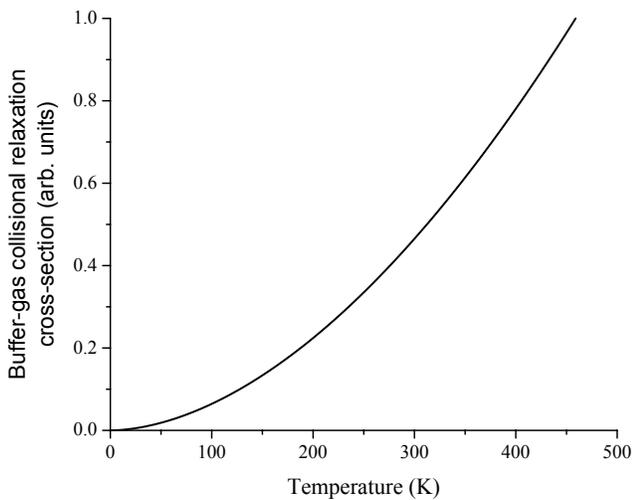}
    \caption{The estimated scaling of Rb-He spin-relaxation cross-section
    with temperature.}
    \label{fig:CrossSectionBG}
\end{figure}

There are, however, compelling theoretical reasons to believe that
buffer-gas collisional relaxation cross-sections should drop quickly
as the temperature is lowered. This relaxation occurs due to the
spin-rotation coupling (which arises from spin-orbit interaction
~\cite{Wu1985,Walker1997,Dashevskaya1971}):
\begin{equation}
\label{eq:SpinRotation} H = \gamma (R)\bm{S}\cdot\bm{N},
\end{equation}
where $\gamma$ is the interaction strength, $R$ is the distance
between the colliding alkali and buffer-gas atoms, $\bm{S}$ is the
alkali electron spin, and $\bm{N}$ is the rotational angular
momentum of their relative motion. In the present work, helium is
used as the buffer gas at cryogenic temperatures, and several other
buffer gases are used at higher temperatures also. Formation of Van
der Waals molecules is neglected~\cite{Walker1989}, and the
spin-relaxation cross-section can be evaluated in the semi-classical
binary collision approximation:
\begin{equation}
\label{equ:BinaryCollision} \sigma(E) = \frac{8\pi
M^2}{3\hbar^4}\int_0^{\infty} b^3 db \left|
\int_{r_0}^{\infty}\frac{\gamma(R)
dR}{\sqrt{(1-b^2/R^2)-V(R)/E}}\right| ^2,
\end{equation}
where $V(R)$ is the inter-atomic potential, $b$ is the impact
parameter, $E$ is the collision energy, $r_0$ is the distance of
closest approach, and $M$ is the reduced mass of the colliding
atoms. The key input into this expression is the R-dependence of
$\gamma(R)$. It is possible to estimate $\gamma(R)$ by approximating
the alkali valence electron wavefunction at large distances $r$ from
the core by the hydrogenic expression (in atomic units)
\begin{equation}
\label{equ:Wavefn} \phi(r)\propto r^{\nu-1}e^{-r/\nu},
\end{equation}
where $\nu$ is the effective principal quantum number of the
ground-state valence electron. Using the results of
Ref.~\cite{Walker1997}, $\gamma(R)$ can be estimated with
logarithmic precision, resulting in the exponential R-dependence:
\begin{equation}
\label{equ:gamma} \gamma(R)\propto e^{-2R/\nu}.
\end{equation}
This is in good agreement with the results of numerical calculations
presented in Ref.~\cite{Walker1997}. It is already possible to argue
why the spin-relaxation cross-section decreases rapidly at low
temperatures. As the temperature is lowered, the collision energy
decreases, and the alkali-helium closest approach distance increases
due to repulsion between the He and the valence electron. Since the
spin-rotation interaction of Eq.~(\ref{equ:gamma}) drops
exponentially, the corresponding spin-flip cross-section also
decreases. We make no attempt at calculating the magnitude of the
relaxation cross-section, instead we estimate its temperature
dependence, using Eq.~(\ref{equ:gamma}) and the interatomic
potential of the form
\begin{equation}
\label{equ:V} V(R) = A R^{2\nu-2}e^{-2R/\nu} - \frac{C_6}{R^6},
\end{equation}
where the first (repulsion) term is taken to be proportional to the
probability density of the alkali electron at the location of the
helium atom, and the second term is the Van der Waals attraction.
The Van der Waals coefficient $C_6$ is known from experiment, and
the parameter $A$ fitted to reproduce the experimentally observed
potential minimum. The result of numerical integration of
Eq.~(\ref{equ:BinaryCollision}) is plotted in
Fig.~\ref{fig:CrossSectionBG} for Rb-He collisions. The interatomic
potential parameters are obtained from
Refs.~\cite{Zhu2004,Herzberg1950}

If indeed, as shown in Fig.~\ref{fig:CrossSectionBG}, the buffer-gas
collisional relaxation cross-section drops rapidly with decreasing
temperature, extremely long ground-state electron spin-relaxation
times can be achieved in alkali atoms inside high-density helium
buffer gas at low temperature. There are, of course, other
spin-relaxation mechanisms that will limit the achievable relaxation
times, such as spin-destruction and spin-exchange in collisions
between the alkali atoms themselves. It is much harder to estimate
the behavior of the corresponding cross-sections at low
temperatures, and the region between 4~K and 300~K remains
experimentally unexplored. Data are available at higher temperatures
for a variety of other buffer gases (such as N$_2$, $^3$He, Xe), a
discussion of the the relevant temperature dependences is given in
Ref.~\cite{Chen2007}. We note, however, that our simple model is not
applicable in these cases, where other effects, such as formation of
Van der Waals molecules, contribute to the spin-destruction
cross-section.

As mentioned above, the reason for scarcity of experimental data
below room temperature is the vanishing saturated vapor pressure of
the alkalis. There has, however, been some recent experimental
effort in this direction. One technique that has been used to study
alkali atoms in helium-filled cells at cryogenic conditions, is
loading by light-induced atomic desorption (LIAD) from the
liquid-helium film that covers the cell walls below the helium
superfluid-transition
temperature~\cite{Hatakeyama2000,Hatakeyama2002}. These experiments
were performed at 1.85~K, and spin-relaxation times of minutes and
even longer can be inferred from the data. This already supports the
theoretical estimates outlined above. Another technique that has
recently been well developed is production of cold atomic and
molecular beams by laser ablation inside helium buffer gas at
temperatures on the order of 5~K~\cite{Maxwell2005}. The same
technique has also been used for loading atoms and molecules into
magnetic traps~\cite{Kim1997,deCarvalho1999}. The experiments
described in the present paper fall somewhere in the middle between
these techniques. We use laser ablation to produce high densities of
atomic species inside helium buffer gas in a range of temperatures
between 20~K and 300~K, but, in addition to cooling the ablated
atoms, the buffer gas also confines them by slowing down transport
to the walls, where the atoms condense.

\section{Apparatus}

\begin{figure}[h]
    \includegraphics[width=\columnwidth]{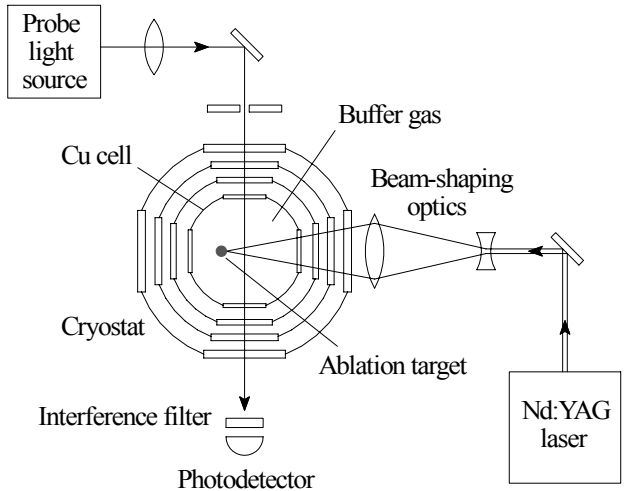}
    \caption{A schematic top view of the experimental setup.} \label{fig:ColdMagnetometerSetUp}
\end{figure}
The experimental setup is schematically shown in
Fig.~\ref{fig:ColdMagnetometerSetUp}. We use a Janis model DT
SuperVariTemp pumped helium cryostat. Optical access is provided via
fused quartz windows, the innermost windows having the diameter of
$1''$. A $2.5''$-tall cylindrical copper cell, with inner diameter
of $1.25''$ is mounted inside the cryostat, in line with the
windows. The cell prevents the ablated species from being deposited
on the inside walls and windows of the cryostat sample space, which
are hard to clean. The copper cell has four $1''$-diameter fused
quartz windows, and several holes at the top and bottom to make sure
that the buffer gas pressure inside the cell is equal to that inside
the cryostat sample space. The ablation target is mounted on a
holder inside the cell. At the top of the cryostat, the sample space
is connected to a Stokes Pennwalt rotary vane pump via a
$3/4''$-diameter pumping arm. Also at the top, there is a connection
to an MKS Baratron Type 220B pressure gauge and a valve for letting
in buffer gas from a pressurized cylinder (99.995\% purity). The
temperature is monitored with a LakeShore silicon-diode temperature
sensor, model DT-470-CU-13-1.4L, mounted on the top of the copper
cell. LakeShore model 321 controller, combined with a wound heater
mounted next to the temperature sensor, allows temperature control
during the experiment.

A Q-switched Spectra-Physics DCR-11 Nd:YAG laser, operating at the
fundamental wavelength (1064~nm), is used to ablate the targets. The
laser beam passes through a set of lenses to expand it (this is done
to avoid damaging the cryostat windows by the focused beam), and is
then focused to a spot on the ablation target by a
large-numerical-aperture lens. The focus (the ablation spot) can be
moved around the surface of the target by moving this lens, which is
mounted on a three-coordinate translation stage. Several different
ablation-laser operating regimes have been explored. It was found
that the ablation yield is independent of pulse energy when it is
above 30~mJ, but drops when the pulse energy is lower. Doubling the
frequency of the ablation pulse and operating at 532~nm had no
noticeable effect on the ablation yield. Firing the laser with the
Q-switch disabled gave a ``long'' 100-$\mu$s light pulse. When such
a pulse was used for ablation, the resulting atomic yield was at
least an order of magnitude lower than that obtained with the
Q-switched laser. After exploring these ablation regimes, the
following parameters were chosen for data taking: pulse energy of
50~mJ, fundamental wavelength (1064~nm), Q-switched (10-ns light
pulse).

\section{Silver and gold}

Four atomic species have been studied in our experiment: lithium,
rubidium, silver and gold. We start with the discussion of ablation
of silver and gold. These atoms were chosen, on the one hand,
because of their ground-state electronic configurations, consisting
of a single s electron outside a filled shell. They have S$_{1/2}$
ground states, and their spectra are similar to those of alkali
atoms. On the other hand, unlike the alkalis, silver and gold are
ductile, and thin wires (diameter as small as $25\mum$) are readily
available. We decided to use such thin wires as ablation targets for
the following reasons. The amount of material that has to be
vaporized to create an atomic density of $10^{11}\invcucm$ inside
the cell of volume 20~cm$^3$ is only $2\times10^{12}$~atoms. The
energy required to evaporate that amount of material is
approximately 1~$\mu$J. The rest of the energy delivered to the
target by the ablation pulse goes into the buffer-gas shock wave,
ejects macroscopic pieces of the target, or is conducted away into
the bulk of the target, heating it up, and heating the buffer gas
with it. Using the tip of a sub-millimeter diameter wire as the
ablation target allows the heat conduction away from the ablation
spot to be minimized, because the wire's cross-sectional area
through which the heat escapes is small. In addition, the thin wire
has a much smaller surface area than a flat target, so the
probability of ablated atoms to diffuse back and stick to the target
surface is minimized.

Silver and gold wires of 50~$\mu$m diameter (purchased from Alfa
Aesar) were used as ablation targets. Measurements were also taken
using wires of 25~$\mu$m and 100~$\mu$m diameter; wire thickness did
not significantly affect experimental results in this range. The
wire was mounted vertically in the middle of the cell inside the
cryostat sample space, so that the tip of the wire was just below
the center of the cell windows. Aiming the ablation laser focus at
the tip of the wire was done as follows. The laser intensity was
reduced by a neutral density filter to avoid ablation. When the wire
was in the focus, it blocked part of the beam, and a shadow was
clearly visible on the beam stop, where the laser beam was projected
after it exited the cryostat. Once the laser was aimed directly at
the wire tip, the neutral density filter was removed and an ablation
shot was fired. As the material at the wire tip was evaporated, the
ablation spot was moved up by translating the focusing lens.

Ablated atoms were detected by measuring the absorption of a light
beam resonant with the D1 atomic transition. This light beam was
produced by a hollow cathode lamp (HCL) and focused with a lens. A
narrow-band interference filter was used to select the D1 atomic
line. For the silver HCL a 338-nm filter was used, and for the gold
HCL a 270-nm filter was used. In each case the transmitted spectrum
was measured with a Czerny-Turner grating spectrometer, confirming
that only a single spectral line was present. We also checked that
the observed signal was indeed resonant atomic absorption rather
than some broad-band scattering: when a helium-neon laser beam was
used as the probe in place of the HCL, no absorption was detected.

\begin{figure}
    \includegraphics[width=\columnwidth]{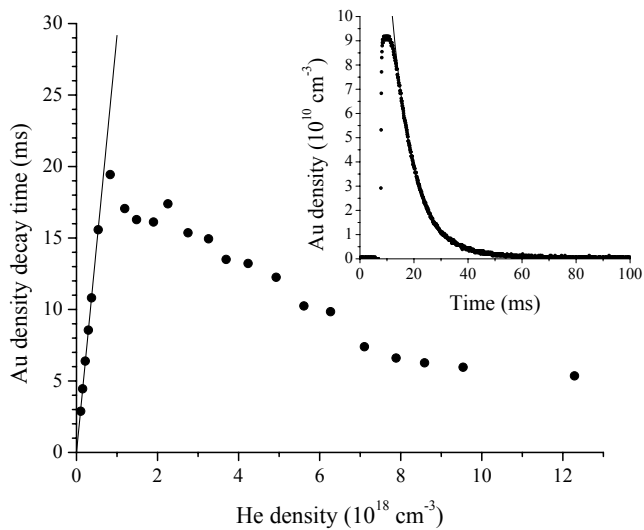}
    \caption{Gold-atom density decay time as a function of helium density at
    295~K. The straight line is a linear fit to the low-helium-density data,
    where diffusion dominates the gold atom loss.
    The inset shows the evolution of gold-atom vapor density after a single ablation shot,
    for helium density of $7\times 10^{18}\invcucm$. The line is an exponential fit to
    the density decay.} \label{fig:AuHe}
\end{figure}
The inset in Fig.~\ref{fig:AuHe} shows the time-evolution of
gold-atom density after a single ablation shot is fired into the tip
of the gold wire. Gold atoms fill the cell after the ablation laser
pulse. The photo-detector bandwidth is not wide enough to make an
accurate measurement of the filling time. This time is faster than
100~$\mu$s, and, in the entire experimental range of buffer gas
densities, is always much faster than the time it would take for Au
atoms to diffuse from the wire to the location of the probe beam,
which is 5~mm away. The filling process must be non-diffusive $-$ a
shock wave propagates in the helium gas after the ablation pulse,
filling the cell with ablated Au atoms. Typical Au densities
achieved are on the order of $10^{11}$~cm$^{-3}$. After a maximum is
reached, the atomic density starts to decrease. Exponential decay
fits the data well, and the loss of Au atoms from the buffer gas can
be characterized by the corresponding decay time. The mechanisms
causing this Au atom density decay are discussed below.

The evolution of Au atom density after an ablation shot at each
helium density and room temperature were recorded, and the resulting
decay times are plotted in Fig.~\ref{fig:AuHe}. At small helium
density, the Au loss time grows linearly with helium density. In
this regime, Au loss due to diffusion to the cell walls dominates.
The atoms are lost at the walls since the temperature is too low to
support any significant Au vapor pressure. An estimate for the
diffusion time $t_d$ can be written as:
\begin{equation}
\label{equ:Diff1} t_d \simeq \frac{R^2}{6D},
\end{equation}
where $R\simeq1.5$~cm is the distance from the probe beam to the
cell walls, and $D$ is the diffusion coefficient: $D\simeq\lambda
v/3\simeq v/3\sigma_t n_{\text{He}}$. Here $\lambda$ is the mean
free path of Au atoms, $\sigma_t$ is their transport cross-section,
$v$ is the rms relative velocity, and $n_{\text{He}}$ is the helium
density. The diffusion time can be expressed as
\begin{equation}
\label{equ:Diff2} t_d\simeq \frac{R^2\sigma_t}{2v}n_{\text{He}}.
\end{equation}
Indeed the diffusion time grows linearly with increasing helium
density, and the diffusion coefficient $D_{\text{AuHe}}$ for gold
atoms in helium can be extracted from the slope. The result for the
temperature of $295\K$ and atmospheric pressure is:
$D_{\text{AuHe}}\simeq0.5\sqcm$/s (which corresponds to a reasonable
value of the transport cross-section, $\sigma_t\approx
10^{-15}$~cm$^2$). No published value for this diffusion coefficient
has been found in literature.

If diffusion to the walls were the only loss mechanism, Au vapor
lifetime would continue to grow linearly with increasing helium
density, and reach about a second at the density of $\approx 3\times
10^{19}$~cm$^{-3}$ . As evident from Fig.~\ref{fig:AuHe}, however,
at helium density of about $10^{18}\invcucm$ another Au loss
mechanism starts to dominate, and the Au vapor lifetime decreases
with increasing helium pressure. One candidate for this loss is
dimer formation: Au + Au $\rightarrow$ Au$_2$. When such a dimer is
formed, the Au atoms are lost, since they are no longer resonant
with the probe laser. A collision between two gold atoms in vacuum
can not form a dimer, since both energy and momentum can not be
conserved in such a process. A third body is needed, and a helium
atom can fulfill this role if it is close enough to the colliding Au
atoms. Dimer formation rate then grows with growing helium density,
and this can explain the decrease of gold lifetime as helium density
increases. Simple estimates predict dimer formation rates that are
about an order of magnitude too small to explain the observed gold
atom loss, but more accurate calculations are needed to rule out
this mechanism.

Another possible gold atom loss mechanism is capture by clusters:
Au~+~Au$_n$~$\rightarrow$~Au$_{n+1}$. For this process there is no
need to have a helium atom nearby, since excess kinetic energy can
be converted into vibrational energy of the cluster and subsequently
carried away by collisions of the cluster with helium atoms. Hence
there is no dependence of the cross-section on the helium density.
The interaction between a gold atom and a cluster is well described
by the Van der Waals potential, and, as a consequence of high
cluster polarizabilities, the cross-sections are extremely large, on
the order of $10^{-13}\sqcm$, for clusters with
$n\simeq10$~\cite{Kresin1998}. If we take this value as the
gold-atom capture cross-section, the number density of clusters
required to produce gold atom loss on the time scale of 10~ms is
$10^{11}\invcucm$. Where would these clusters come from? As argued
previously, dimer formation is too slow to lead to nucleation of an
appreciable number of clusters. The ablation process itself,
however, is likely to produce such
clusters~\cite{Nichols2002,Anisimov2002}. As the ablation laser hits
the target, it locally heats the material, creating an explosion.
The explosion forms a shock wave, which expands out into the helium
gas. Some of the high-density gold vapor created by the explosion
condenses into clusters in the low-pressure region behind the shock.
The higher the helium density, the greater the efficiency of this
condensation, the more clusters are produced, and the faster the
rate of gold vapor loss. This may explain the trend observed in
Fig.~\ref{fig:AuHe}.

The third possible loss mechanism is capture of a gold atom by an
impurity (such as a hydrocarbon or an oxide molecule) that is either
present in the buffer gas to start with, or gets ablated off the
target surface. Such an impurity can easily have a large capture
cross-section, and the impurity density of $10^{11}\invcucm$ or more
is not at all implausible.

\begin{figure}
    \includegraphics[width=\columnwidth]{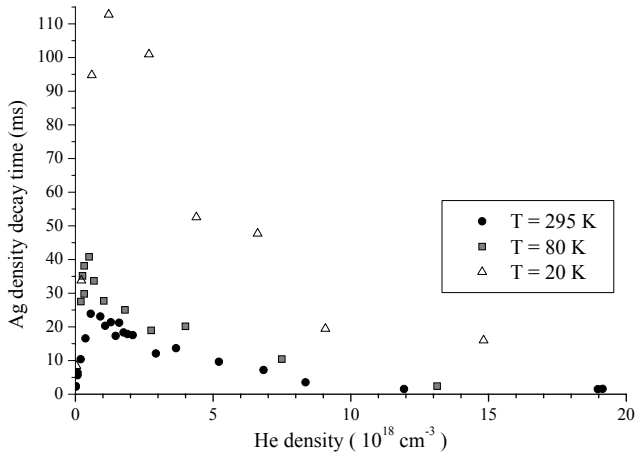}
    \caption{Silver vapor density decay time as a function of helium buffer gas
    density at different temperatures.} \label{fig:AgHe}
\end{figure}
A study of the temperature dependence of the atomic vapor lifetime
was performed using silver atoms. The silver vapor-density decay
times are plotted in Fig.~\ref{fig:AgHe} for laser ablation of a
$50\mum$-diameter silver wire in helium buffer gas at three
temperatures. The trends are similar to the ones in
Fig.~\ref{fig:AuHe}, where gold vapor results are shown. At small
helium density, the silver vapor lifetime grows linearly, and the
main atom loss mechanism is diffusion to the cell walls. The
diffusion coefficient $D_{\text{AgHe}}$ for silver atoms in helium
can be extracted from the slope of the linear dependence of
diffusion time on helium density, using Eq.~(\ref{equ:Diff1}). The
result for the temperature of $295\K$ and atmospheric pressure is:
$D_{\text{AgHe}}\simeq0.4\sqcm$/s. Silver vapor lifetime reaches a
maximum at $n_{\text{He}}\simeq10^{18}\invcucm$ and decreases for
larger helium densities. The silver atom loss mechanisms are likely
the same as those for gold, described above. It can be seen in
Fig.~\ref{fig:AgHe} that at low temperatures, longer atomic vapor
lifetimes can be achieved. The longest observed silver vapor
lifetime was 113~ms at the temperature of 20~K~\cite{NoteAg}. The
peak silver atomic densities were on the order of $10^{11}\invcucm$.
Data were also taken at 80~K and 295~K with three other buffer
gases: nitrogen, neon, and argon. It was found that the atomic-vapor
decay time at a given buffer gas density does not noticeably depend
on which of these buffer gases is used.

\section{Lithium and rubidium}

Studying silver and gold atoms is difficult because their atomic
transition frequencies lie in the UV: the D1 lines are at the
wavelengths of 338~nm for silver and 268~nm for gold. The
hollow-cathode lamps are sufficient for absorption measurements, but
for optical pumping a more intense light source is needed. Such
light sources (diode lasers) are readily available for alkali atoms.
We therefore undertook a study of alkali-atom ablation with our
experimental setup. Experiments with lithium and rubidium were
performed. For lithium the ablation target was in the form of
0.75-mm-thick and 19-mm-wide lithium foil (purchased from Alfa
Aesar), and for rubidium the target was a roughly 5-mm-thick and
10-mm-wide piece of pure rubidium metal. Alkali metals are extremely
reactive, and they oxidize quickly when exposed to air. Therefore,
ablation targets were prepared in a glove box with argon atmosphere.
The target was mounted inside the copper cell, level with the cell
windows. The glove box was then opened and the cell was quickly
inserted into the cryostat, which was immediately pumped out. In
this way the exposure of the alkali targets to air was minimized.

\begin{figure}[b]
    \includegraphics[width=\columnwidth]{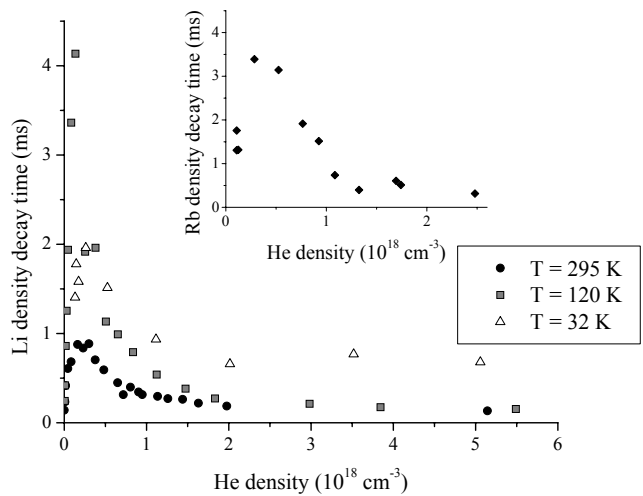}
    \caption{Lithium vapor-density decay time as a function of helium buffer gas
    density at different temperatures. The inset shows rubidium data at 295~K.} \label{fig:LiHe}
\end{figure}
Ablated atoms were detected by measuring the absorption of a laser
beam resonant with one of the D1 or D2 atomic transitions. Lithium
atoms were probed with the light from a 671-nm laser diode (TOLD
9221M) inside a temperature-controlled mount (Thorlabs TCLDM9)
paired with a TED-200 temperature controller and a LDC-201-ULN
current controller. A 750-MHz free spectral range confocal
Fabry-Perot spectrum analyzer (F-P) was used for spectral
diagnostics of the laser light. Tuning to the $^7$Li (92.5\%
abundance) D1 resonance line at 671~nm was achieved by splitting off
a part of the beam, modulating the intensity with a chopper wheel at
350~Hz, and aiming it into the cathode of a lithium hollow-cathode
lamp (HCL). When the laser was tuned to the lithium resonance, the
HCL current was modulated at the chopper frequency (opto-galvanic
effect)~\cite{King1978}. This modulation was detected with a lock-in
amplifier, referenced to the chopper-wheel frequency. The setup for
rubidium atoms was similar, with the 671-nm laser diode replaced
with a 780-nm laser diode, and the hollow cathode lamp replaced with
a rubidium vapor cell, which was used to tune the laser to the Rb D2
transition by detecting atomic fluorescence.

The time-evolution of lithium and rubidium atom density after an
ablation shot is very similar to that shown on the inset in
Fig.~\ref{fig:AuHe} for gold. The ablated atoms fill the cell very
quickly after the laser pulse, a maximum is reached, and then the
atomic density decreases. Exponential decay fits the data well, and
the decay times are plotted in Fig.~\ref{fig:LiHe} for lithium in a
range of temperatures. Rubidium data at room temperature are shown
in the inset. The trends are the same as those seen for gold and
silver atoms. At low helium density the vapor loss time increases
linearly, in agreement with Eq.~(\ref{equ:Diff2}), which describes
diffusion to the walls. But the maximum in the decay time now occurs
at lower helium density, of approximately
$0.2\times10^{18}\invcucm$. At higher densities another loss
mechanism dominates, keeping the decay times short. As discussed
above, the possible candidates for this mechanism are dimer
formation, atom loss on sputtered clusters, and atom loss on
impurities in the buffer gas. For an unknown reason, this mechanism
is more efficient for alkali atoms than for silver and
gold~\cite{NoteWireD}. This leads to shorter vapor-loss times for
the alkalis, on the order of 4~ms at the maximum. Data were also
taken at 295~K with nitrogen as the buffer gas. It was found that
the lithium vapor decay time at a given buffer gas density does not
noticeably depend on whether the buffer gas is helium or nitrogen.

\section{Conclusions and outlook}

In conclusion, we have used laser ablation to create and study
atomic vapors of silver, gold, lithium, and rubidium inside helium
and other buffer gases in a wide range of temperatures and
buffer-gas pressures. Vapor densities of $10^{11}\invcucm$ have been
achieved. The longest measured atomic vapor lifetimes are on the
order of $110\ms$ for silver~\cite{NoteAg} and $4\ms$ for lithium.
For buffer-gas density $n\ll 10^{18}\invcucm$ the atomic vapor loss
is dominated by diffusion to the cell walls, where the atoms
condense. Increasing the buffer gas density above $10^{18}\invcucm$,
however, does not increase the atomic vapor lifetime, instead the
lifetime gets shorter. The possible loss mechanisms that can cause
this behavior are dimer formation, atom loss on clusters created
during the ablation process, or atom loss on impurities in the
buffer gas.

To put our results in context, let us consider the figure of merit
of an atomic magnetometer, or any shot-noise limited precision
measurement with an ensemble of $N$ atoms with coherence time
$\tau$~\cite{Auzinsh2004,Budker2007}:
\begin{equation}
\label{equ:FOM} \text{FOM}\propto\sqrt{N\tau}.
\end{equation}
The coherence times have not been measured in our experiments so far
(these measurements are currently in progress). However, as argued
at the beginning of this paper, the buffer-gas collisional
relaxation rates at cryogenic temperatures are expected to be on the
order of millihertz. If atom loss as measured in our experiments
proves to be the dominant mechanism for coherence relaxation, then
coherence relaxation rates on the order of several hertz are
expected. With atomic vapor densities of $10^{11}\invcucm$ already
demonstrated, the figure of merit for our system is competitive with
the leading atomic magnetometers today~\cite{Budker2007}. Our setup,
however, has the advantage of operation in a wide range of
temperatures: from 4~K to 300~K, while the vapor-cell based
measurements can only be performed at room temperature and above.


The authors thank Valeriy Yashchuk and Max Zolotorev for many useful
discussions. This research has been supported by the National
Science Foundation through grant 0554813.

\bibliography{BufferGas_ArXiV_v2}

\begin{thebibliography}{32}
\bibitem[{Not({\natexlab{a}})}]{NoteAg}
\bibinfo{note}{After this article was completed, measurements at slightly
  different experimental conditions (lower temperatures, for example) showed
  even longer silver atom lifetimes, up to 300 ms.}
\bibitem[{Not({\natexlab{b}})}]{NoteWireD}
\bibinfo{note}{This may be due to the fact that only bulk targets were used for
  alkali ablation, while sub-millimeter diameter wires were the targets for
  silver and gold. On the other hand, as mentioned in Section III, the wire
  diameter (in the range 25 to 100 microns) did not noticeably affect
  experimental results for silver and gold.}
\end{thebibliography}

\end{document}